# Virtual Inertia Control of the Virtual Synchronous Generator: A Review

M. Li, W. Huang, N. Tai, D. Duan

*Abstract*— **With the increasing impact of low inertia due to the high penetration of distributed generation, virtual synchronous generator (VSG) technology has been proposed to improve the stability of the inverter-interfaced distributed generator by providing "virtual inertia". This paper presents a recent review of virtual inertia control covering significance, features, design principles, and state-of-art inertia strategies from both physical and mathematical perspectives to facilitate the wide application of the VSG. The definition and source of virtual inertia are given to help researchers to establish the concept of "virtual inertia". Then, this paper covers influencing mechanism studies of virtual inertia to reveal its functions. Also, a design framework of the virtual inertia is established by considering both the characteristics of the control system and the limitation of energy storage systems and renewable energy resources. Finally, several novel adaptive inertia control strategies are reviewed, and some aspects of potential future research are recommended.**

## I. Introduction

THE integration of distributed energy resources in the power system is increasing rapidly all over the world [1, 2]. Distributed generation using renewable energy resources, battery energy storage systems, super-capacitor energy storage, etc. is based on fast-response inverters, which decreases power system inertia and brings challenges to the stable operation [3-6].

In order to address these problems, the control scheme of the virtual synchronous generator (VSG) for inverter-interfaced distributed generators is proposed by incorporating the swing equation of synchronous generators to the control system [7]. In this scheme, the system is provided with additional inertia during transient disturbances through storage units. This virtual inertia emulation could improve frequency response and increase the stability of power systems. Besides, since most VSG controllers are based on the principles of droop controller [8], the VSG control has the advantages of droop control, such as power-sharing among distributed generators, plug-and-play support, and flexible transition between the islanded and the grid-connected mode. The VSG technology is increasing in importance as a solution to large-scale integration of distributed generators. There have been some notable demonstration projects of the application of VSG technology. For instance, the wind power-photovoltaic generation-energy storage system in Zhangbei County, Hebei, China, is the largest VSG demonstration project of photovoltaic generation around the world, where the accumulated installed capacity of VSG reaches to 140 MW [2]. The VSG based photovoltaic system-energy storage system in Wuxi County, Chongqing, China, provides an example for typical power systems of remote mountain areas [6].

Virtual inertia constant is the core parameter influencing the performance of the VSG. It provides an inertia property that distinguishes the VSG from other control schemes, especially the classic droop control [9, 10]. From the perspective of the mathematical governing equation, the time scale and dynamic characteristics of a VSG control system depend on the range of the inertia [11]. Compared to the rotating inertia of a real synchronous generator, which is fixed when the machine is made, the virtual inertia constant is a control parameter allowing for greater flexibility. Smaller or larger values than that of a real synchronous generator are permitted, and it can be adjusted dynamically to meet with different operational requirements. As for the perspective of physical realization, the virtual inertia emulation relies on the energy storage device, like the battery, super-capacitor, flywheel, rotors, etc. These energy storage units are required to absorb and release energy to decrease power fluctuations following disturbances.

In the past decade, large numbers of works have been published proposing new inertia control strategies [4, 9, 13-17] to improve the stability of a VSG system. Mathematic models

This work was supported by the No. 51807117, National Natural Science Foundation of China and No. 201901070002E00044, Major Scientific Research and Innovation Project of Shanghai Education.

M. Li is with the Department of Electrical Engineering, Shanghai Jiao Tong University, Shanghai 200240, China (e-mail: my546277681@163.com).
W. Huang is with the Department of Electrical Engineering, Shanghai Jiao Tong University, Shanghai 200240, China (Corresponding author, e-mail: hwt8989@sjtu.edu.cn).
N. Tai is with the Department of Power Electrical Engineering, Shanghai Jiao Tong University, Shanghai 200030, China (e-mail: nltai@sjtu.edu.cn).
D. Duan is with the Department of Electrical and Computer Engineering, University of Wyoming, WY 82071 USA (e-mail: dduan@uwyo.edu).
M. Yu is with the Department of Electrical Engineering, Shanghai Jiao Tong University, Shanghai 200240, China (e-mail: m18817519493@163.com).

have been established in [18-21] to analyze the effect of virtual inertia. As the VSG control scheme becomes a heated issue, a few reviews of the VSG have been reported, as shown in Table I., The research structure in [2] divides the VSG studies into five parts: underlying control, modeling, VSG algorism, stability analysis, and application. Reference [22] presents the main application of the VSG with the combination of photovoltaic (PV) generation, wind power, energy storage system, and the load. Reference [3, 23] summarizes the main methods of stability analysis of the VSG. These studies mainly focus on motivation, control structure, dynamic characteristics, and features of the VSG control system and its stability in the microgrid or distributed-generators-dominated distribution network. However, a systematic and in-depth framework related to physical significance, source, and realization of virtual inertia emulation remains to be established. With the increasing impact of low inertia on grid stability, there is a need to provide a comprehensive review of the virtual inertia control to help researchers identify the research framework and get a better understanding of the concept of "virtual inertia" in the VSG control scheme.

TABLE I
THE CONTRIBUTIONS OF PREVIOUS REVIEWS

| References | Major contributions |
|---|---|
| [2] | ☑ Proposed research structure dividing the VSG studies into: underlying control, modeling, VSG algorism, stability analysis, and application |
| [22] | ☑ Main application of the VSG with the combination of different types of distributed generators |
| [3][23] | ☑ Main methods of stability analysis of the VSG |

This paper contributes with a comprehensive review of the virtual inertia control of a VSG that starts with definition, sources, mechanisms, moving to design, utilization, and potential future developments of the virtual inertia control. As shown in Fig.1, this paper provides answers to these questions: What is virtual inertia? Where does virtual inertia come from? How does virtual inertia work? How much virtual inertia can we use? And how to control virtual inertia?

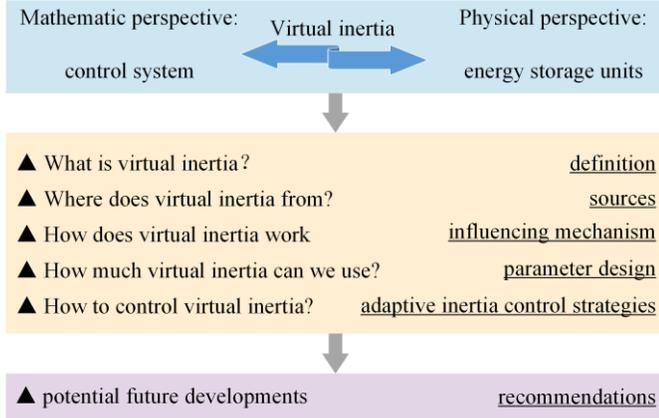

Fig. 1. Research Framework of virtual inertia control

The paper is organized as follows. The definition of virtual inertia constant is introduced in section II. Sources of virtual inertia are summarized in section III. Section IV discusses the influencing mechanism studies of the virtual inertia. Section V presents a framework of the virtual inertia design. Section VI lists several adaptive inertia control strategies to enhance the stability of a VSG. Finally, section VII provides a list of recommendations for further research on the VSG.

II. DEFINITION OF VIRTUAL INERTIA CONSTANT OF THE VSG

A. *VSG Control*

The VSG topology is shown in Fig. 2. By governing the output of the inverters locating between a DC bus and the grid, a VSG has the dynamic features of a conventional synchronous generator from the grid point of view. The swing equation of conventional synchronous generators is introduced as the core of the algorithm:

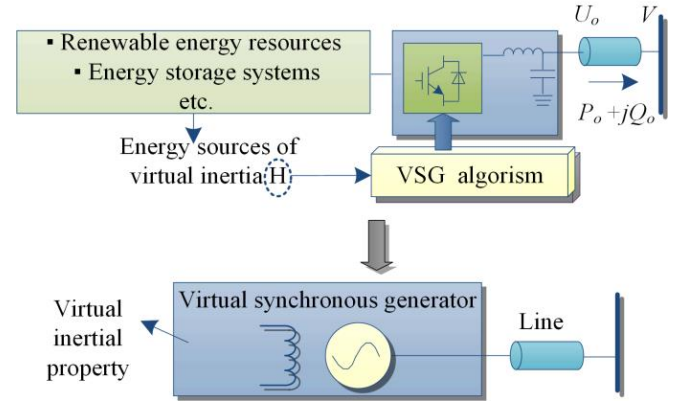

Fig. 2. VSG topology

$$J\dot{\omega} = T_m - T_e - T_D \tag{1}$$

where $J$ is the moment of inertia, $T_m = P_{ref}/\omega$ is the mechanical torque, $\omega$ is the angular frequency, $T_e = P_o/\omega$ is the electromagnetic torque, and $T_D$ is the damping torque.

Standardize (1), the governing equation of the VSG is as (2):

$$H\frac{d\omega}{dt} = P_{ref} - P_o - P_D \tag{2}$$

where $P_{ref}$ is the reference of active power, $P_o$ is the active power output, $P_D = k(\omega - \omega_{ref})$, $k$ is the damping factor, and $\omega_{ref}$ is the reference of angular frequency. $H$ is the virtual inertia constant.

Equation (1) and (2) are two major types of VSG's frequency controller. Equation (1) uses virtual angular frequency $\omega$ to calculate the torque, but equation (2) assumes the angular frequency is $\omega_{ref}$. An extra feedback loop is needed by using equation (1). For unification, the detailed derivation below is based on the equation (2), and the virtual inertia constant is represented by $H$ and the damping factor by $k$. However, it should be pointed out that the effects of virtual inertia in a VSG using equation (1) or (2) are similar.

B. *Definition of the Virtual Inertia H*

From the perspective of dynamic characteristics of the active power control loop, reference [25] defined the virtual inertia constant. When there is no load ($P_o = P_D = 0$), integrate both sides of equation (2) with rated power reference ($P_{ref} = 1$). When the output reaches the rated angular frequency ($\omega = \omega_{ref}$), we could obtain:

$$\int_0^1 H d\omega = \int_0^t (P_{ref} - P_o - P_D) d\tau \Rightarrow H = t \qquad (3)$$

Therefore, the virtual inertia constant is the needed time that the output of a VSG with no load reaches the rated angular frequency with rated power reference.

Reference [24] also regards the virtual inertia as a time constant of the power flow as below. Virtual inertial property is emulated by controlling the active power from the energy storage units through the inverter to the grid in a specific relationship with the rotor speed.

$$H \omega_g \frac{d\omega_g}{dt} = \Delta P_g \qquad (4)$$

Here $\omega_g$ is the grid frequency, $\Delta P_g$ is the power flowing from the grid into the load that produces a frequency deviation.

Many studies analyze the function of the virtual inertia by comparing classical droop control and VSG control [9, 10]. The comparison of frequency governing equations of the classical droop control and the VSG control is:

$$\begin{cases} H \dfrac{d\omega}{dt} = P_{ref} - P_o - k(\omega - \omega_{ref}), & \text{VSG control} \\ 0 = P_{ref} - P_o - \dfrac{1}{m_p}(\omega - \omega_{ref}), & \text{droop control} \end{cases} \qquad (5)$$

where $m_p$ is the droop coefficient of active power loop in the droop control scheme. According to (5), the virtual inertia constant serves as an integration constant in the active power loop. Considering that PI control shows higher flexibility than P control, the VSG control achieves better dynamic characteristics than droop control with this integration constant.

Reference [9] compares the droop control with a low-pass filter using a considerable time constant in the active power droop and the VSG control. Considering that the active power $P_o$ is low-pass filtered, the relationship between active power output $P_o$ and measurement feedback signals $P_{meas}$ is:

$$P_{meas} = (1 + sT_f) P_o \qquad (6)$$

where $T_f$ is the time constant of the first-order low-pass filter. Substitute the active power measurement $P_{meas}$ in (2). The governing equations of the VSG control and the droop control can be expressed as below.

$$\begin{cases} H \dfrac{d\omega}{dt} = P_{ref} - P_o - k(\omega - \omega_{ref}), & \text{VSG control} \\ T_f \dfrac{1}{m_p} \dfrac{d\omega}{dt} = P_{ref} - P_o - \dfrac{1}{m_p}(\omega - \omega_{ref}), & \text{droop control (filter)} \end{cases} \qquad (7)$$

As shown in (7), the governing equation of the droop control with a low-pass filter has the same form as the VSG control. The time constant of the low-pass filter $T_f$ offers similar functionality of the virtual inertia constant [24]. Although this comparison is based on many assumptions and simplification of the model, the results provide a new perspective to understand the role of virtual inertia control.

In summary, virtual inertia serves as a time constant of the governor. The concept of "inertia" is related to response time. From this point of view, it is also easy to understand why the inertial property of a traditional droop-controlled or PQ controlled inverter-interfaced distributed generator is not significant. The response time of power electronics is generally very short and could reach a microsecond [31]. Traditional droop control or PQ control doesn't provide extra inertia. With high penetration of distributed generation, inertial property decreases in the power electronics dominated power systems. Most VSGs adopt the damping term $k(\omega - \omega_{ref})$ serving as a droop controller to imitate the output of synchronous generators. Therefore, the VSG control could provide extra inertia to the system.

III. SOURCES OF INERTIA

The main source of inertia of a conventional synchronous generator is the rotating parts. The moment of inertia $J$ of a conventional synchronous generator is a physical quantity related to its size, and its definition is shown below.

$$J = \frac{GD^2}{4} \qquad (8)$$

where $GD^2$ is the moment of the flywheel. Generally speaking, the large size or capability of a conventional synchronous generator leads to large rotor inertia. Large inertia means enhanced frequency stability of the system.

Although there is no such rotating mass in inverters, the virtual inertia in the control system of inverters brings an inertial property. As shown in Fig. 2, the purpose of the VSG control scheme is to provide additional inertia using energy storage units. As shown in Fig. 3, there are two sources of virtual inertia according to the type of energy:

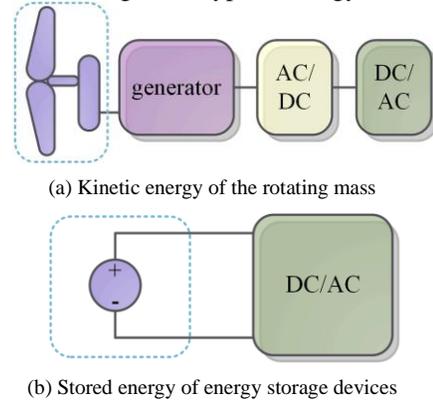

(a) Kinetic energy of the rotating mass

(b) Stored energy of energy storage devices

Fig. 3 Energy sources of the virtual inertia

1) The kinetic energy of the rotating mass, like the rotor of wind turbine generators and tidal turbine generators, as shown in Fig.3 (a);
2) The stored energy of energy storage devices, such as the

battery or DC capacitor, as shown in Fig.3 (b).

The kinetic energy of the rotor is generally limited and challenging to control. The inertia provided by this source is usually to provide instant frequency support for wind turbine generators [4, 21, 30] or tidal power generators [58-59]. By comparison, the virtual inertia offered by energy storage devices is continuous and more flexible. It allows for novel control strategies of the VSG. Many new VSG control strategies have been proposed using the latter form of energy as the source of the virtual inertia, as in [15, 26]. Besides, an energy storage system is usually required to enhance the robustness of a distributed generator (wind generators, photovoltaic units, etc.) against disturbances. Hence, the latter form of energy is more common as the source of the virtual inertia. The two sources of virtual inertia are summarized in Table II.

TABLE II
DIFFERENT SOURCES OF THE VIRTUAL INERTIA

| Type of energy | Kinetic energy | Stored energy |
|---|---|---|
| Energy source | Rotating equipment (wind turbine generators, tidal turbine generators, etc.) | Energy storage devices (battery, capacitor, etc.) |
| Characteristics | ☑Limited ☑Difficult to control ☑Short-term frequency support | ☑Continuous ☑Flexible enough to allow different inertia control strategies ☑More common |

## IV. INFLUENCING MECHANISMS OF VIRTUAL INERTIA

The intrinsic kinetic energy of the real synchronous generator is of great significance in the grid stability of power systems. The inertia of a synchronous generator ensures operation robustness against instability by injecting energy within the proper time interval [32]. Then how does the virtual inertia affect the stable operation of a VSG? It is demonstrated that the virtual inertial constant can contribute to the frequency regulation of a VSG system [8, 16, 33-35]. Several different perspectives of perceiving the functions of virtual inertia constant are discussed below.

### A. Dynamic Characteristics

Responding time and overshot are the most intuitive reflections of dynamic response. Many studies have confirmed that the virtual inertia provides an inertia property which improves the dynamic response [9]. Reference [32] made an analogy of the power-angle curve of a VSG with that of a synchronous generator and showed the virtual inertia influences the responding speed in the acceleration or deceleration progress under disturbances. Some other studies establish mathematical models to analyze the effects of virtual inertia on the overshoot and responding time of the VSG quantitatively, like in [11, 15, 18, 21].

According to (2), the transfer functions of the active power and the angular frequency are as (8) and (9), respectively [11].

$$G_1(s) = \frac{\Delta P_o}{\Delta P_{ref}} = \frac{\lambda}{2Hs^2 + ks + \lambda} \quad (9)$$

$$G_2(s) = \frac{\Delta \omega}{\Delta \omega_{grid}} = \frac{ks}{2Hs^2 + ks + \lambda} \quad (10)$$

where $\lambda = \partial P / \partial \delta |_{\delta=\delta_s, E=E_s}$, $\delta_s$ is the phase angle, $E_s$ is the magnitude of the terminal voltage when the VSG's output reaches reference power.

Fig.4 shows the step response analysis with increasing inertia based on the transfer functions given in equation (9-10). In terms of responding speed, the resettling time of both the angular frequency and active power becomes longer as the virtual inertia increases. As for overshoot, the effects of the virtual inertia on power and frequency are contradictory. As the virtual inertia increases, the overshoot of frequency becomes smaller, but that of power becomes larger. Therefore, the virtual inertia introduces trade-offs to the dynamic responses of a VSG [11], which is summarized in Table III.

TABLE III
DYNAMIC CHARACTERISTICS AS THE VIRTUAL INERTIA INCREASES

| Characteristics | Active power | Angular frequency |
|---|---|---|
| Overshoot | ↑ | ↓ |
| Response speed | ↓ | ↓ |

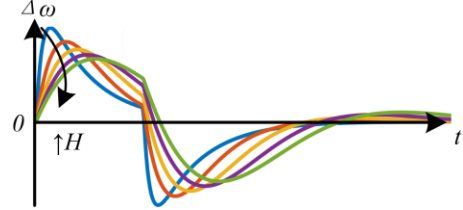
(a) Angular frequency

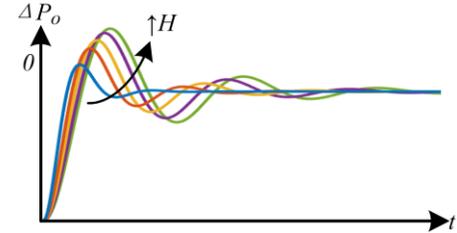
(b) Active power

Fig. 4. Effects of virtual inertia on dynamics responses (Source: [11])

### B. Stability Margin

A disturbance is considered small, such as minor load fluctuations of a VSG system, if a linearized set of equations (a small-signal model) can adequately represent the system behaviour. Otherwise, the large-signal nonlinear analysis should be incorporated [37], such as some faults and unplanned load shedding. Many studies have shown that virtual inertia has dominating effects on the stability margin of a VSG, whether under small disturbances [9, 15, 18, 19, 21, 38] or large disturbances [16, 35].

#### 1) Small-signal Stability

The small-signal stability margin of a VSG is indicated by the distance between the real axis and the eigenvalues.

Eigenvalue analysis based on the state-space matrix or impedance analysis using Nyquist stability criterion or frequency domain characteristics is regular tools to assess the small-signal stability margin. The analysis of power oscillation of a VSG in reference [9] shows that the virtual inertia affects the oscillation frequency, namely the stability margin, but has little effect on the damping ratio. A small-signal model of the current-source VSG with 14 variables is established in [19], and the root locus analysis shows that the dominant poles of the VSG model get closer to the real axis as the virtual inertia increases. Reference [39] focuses on eigenvalue analysis considering the coupling effect between the reactive power loop and the active power loop (APL). The loop gains of the APL with and without the coupling effect of the reactive power loop are as equation (11).

$$\begin{cases} T_P(s) = \dfrac{3V_{on}V_{gn}}{s(Hs+k)X_s}, & \text{without coupling} \\ T_{P,c}(s) = T_P(s)\left(1 - \dfrac{T_q(s)}{1+T_q(s)}\delta_n^2\right), & \text{with coupling} \end{cases} \quad (11)$$

Here $V_{on}$ and $V_{gn}$ are the voltage of the VSG and the grid at the quiescent operating point, respectively. $\delta_n$ is the VSG's angle at the quiescent operating point. $T_q(s) = 3V_{on}D_q k_{iq}/\left(\sqrt{2}D_q X_s(s+D_q k_{iq})\right)$ is loop gain of the reactive power loop without coupling effect. Fig.5 shows the relative root locus plot. The poles of APL with and without the coupling effect of the reactive power loop almost coincide with each other.

Although different control schemes are adopted to establish small-signal models with varying orders, the eigenvalue analysis consistently shows that the small-signal stability margin enhances as the virtual inertia decreases.

*2) Large-signal Stability*

Large-signal stability margin can be obtained by depicting the stability domain of a VSG [14, 40]. Lyapunov-based methods are commonly used in the analysis of large-signal stability. Variations on the inertia of real synchronous generators do not lead to any change of the shape of the stability domain [41]. On the contrary, the shape of the VSG's stability domain changes with different values of inertia. A Lyapunov function is derived in [35] using Popov's stability theory, and the stability domain of a VSG is determined as (12). Here $\theta_s$ is the angle deviation between the VSG and the connected-grid in the equilibrium state. $\theta_\alpha$ is a constant determined by the VSG's configuration.

As shown in Fig. 6, the stability domain of a VSG system becomes smaller as the virtual inertia increases. The shrinking doesn't happen in equal proportion. The oval-shaped stability domain becomes flattered with increasing inertia. However, a large stability domain does not mean enhanced transient stability with longer critical cleaning time [35, 42]. As stated in section IV-A, large inertia is vulnerable to the response speed of power and frequency. It means the distance between an operating point and the stability boundary, or the stability margin, decreases as the virtual inertia increases, and the speed also decreases. Since both the distance and speed decrease, the critical cleaning time of the fault is to be determined [42].

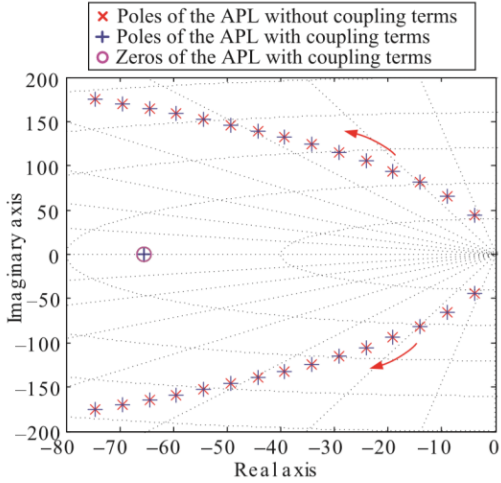

Fig.5 Root locus plot of the APL as the virtual inertia decreases (Source: [39])

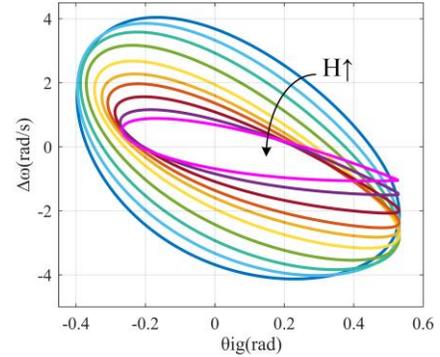

Fig.6 Stability domain with increasing virtual inertia (Source: [35])

$$\text{stability domain}: V(\Delta\theta, \Delta\omega) < M$$

$$\text{where} \begin{cases} V(\Delta\theta,\Delta\omega) = \dfrac{k}{4H}\Delta\theta^2 + \left(\dfrac{H}{k}+\dfrac{k^2}{8H^2}\right)\Delta\omega^2 + \Delta\theta\Delta\omega - \left(\dfrac{\alpha}{k}+\dfrac{\alpha k^2}{8H^3}\right)\left(\cos(\Delta\theta+\theta_s+\theta_\alpha)-\cos(\theta_s+\theta_\alpha)+\Delta\theta\sin(\theta_s+\theta_\alpha)\right) \\ M = \dfrac{(\pi-2(\theta_s+\theta_\alpha))^2}{2}\left(\dfrac{k^4}{32H^4+4Hk^3}\right)-\left(\dfrac{\alpha}{k}+\dfrac{\alpha k^2}{8H^3}\right)\left(\sin(\theta_s+\theta_\alpha)(\pi-2(\theta_s+\theta_\alpha))-2\cos(\theta_s+\theta_\alpha)\right) \end{cases} \quad (12)$$

*C. Transient Energy*

The conventional synchronous generator will inject the kinetic energy preserved in its rotating parts to the grid when it is subject to disturbances. However, inverter-interfaced distributed generators rely on power inverters, which do not possess any intrinsic inertia. It results in inadequate balancing energy injection under disturbances. The VSG control scheme provides solutions to this problem. The analysis in [32] shows that virtual inertia can help to adjust the virtual transient energy under disturbances, and finally, the system will reach the equilibrium point. The VSG's transient energy following a disturbance is derived as below.

$$V = \underbrace{\frac{1}{2}\omega_{ref}H\Delta\omega^2}_{E_k} \underbrace{-\left[P_{in}(\delta-\delta_s)+b(\cos\delta-\cos\delta_s)\right]}_{E_p} \quad (13)$$

Here $b$ is the amplitude of the power-angle curve, $E_k$ and $E_p$ are the virtual kinetic energy and potential energy of the VSG system, respectively.

When the system is subject to a disturbance, $\Delta\omega$ is zero, and $\delta-\delta_1$ is the maximum value. Therefore, $E_k=0$ and $E_p$ is the maximum value. During the recovery, $E_k$ increases as $\Delta\omega$ increases, and $E_p$ decreases as $\delta-\delta_1$ decreases. Or otherwise. The system transient energy is converted between the potential form and the kinetic form during oscillations. The theorem is shown in Fig. 7(a).

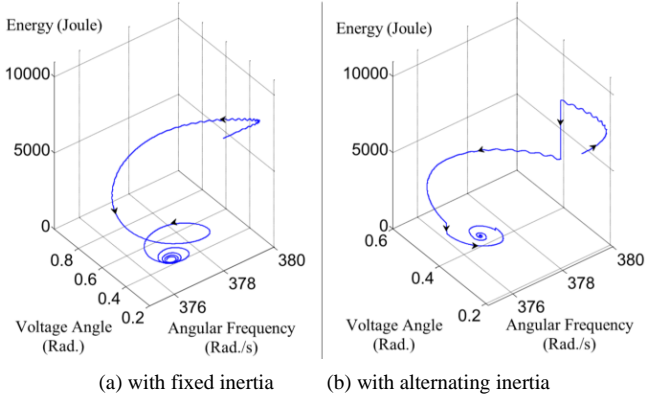

(a) with fixed inertia    (b) with alternating inertia

Fig. 7. Transient energy of a VSG system after disturbances (Source: [32])

According to (13), the virtual inertia affects the energy in the kinetic form and the system transient energy level. In [32], the virtual inertia is adjusted during the oscillation to analyze the damping effect of the virtual inertia on transient energy, and the results are shown in Fig. 7(b). When all system transient energy is converted to the kinetic form, decrease the virtual inertia. The VSG's transient energy drops and the amplitude of oscillation will be reduced. Then, increase the inertia when all the transient energy is converted to potential form. The transient energy remains the same, according to (13).

In summary, the virtual inertia introduces more flexibility in the control system for a VSG. Fig.8 summarizes the effects and function of virtual inertia:

1) By selecting a different value of the virtual inertia, the VSG's dynamic characteristics can be improved with fast response and damped oscillations.
2) Large virtual inertia has a deteriorative effect on the small-signal stability margin. Small virtual inertia leads to extended large-signal stability domain.
3) The virtual inertia is proportional to the energy in the kinetic form, and further affects the system transient energy. Damped oscillations could be obtained if applying the small virtual inertia when all system transient energy is in the kinetic form.

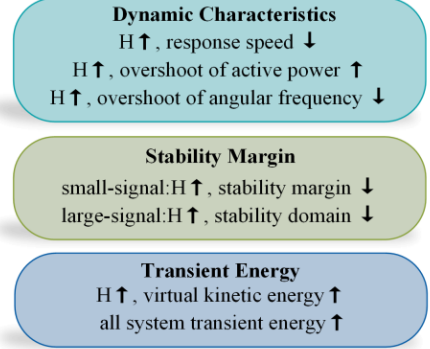

Fig. 8 Effects and function of virtual inertia

## V. DESIGN OF VIRTUAL INERTIA

The parameter selecting method of a VSG is like the other control schemes for inverters, where control theory serves as the main tool. From the perspective of the mathematical governing equation, virtual inertia serves as a time constant of the governor, and the time scale of a VSG control system depends on the range of the inertia [11]. The moment of inertia of conventional synchronous generators is fixed when the machine is made. On the contrary, the inertia control parameter of a VSG is virtual and allows for greater flexibility. Smaller or larger inertia than that of a real synchronous generator is permitted to achieve better dynamic and transient performances of a VSG [43]. The range of virtual inertia is much larger than that of rotating inertia of a real synchronous generator. It can even be adjusted dynamically to meet with different operational requirements.

As for the perspective of physical realization, the virtual inertia emulation relies on the energy storage device. Virtual inertial property is emulated by controlling the active power from the energy storage units through the inverter to the grid in a certain relationship with the rotor speed. Therefore, a certain capacity of energy storage units is needed to achieve the inertia emulation. Besides, the output of the VSG using renewable energy resources is intermittent and random. The generation profile also influences the output.

As shown in Fig.9, the virtual inertia design framework is established in this section. Two aspects are considered to select suitable virtual inertia for the VSG:
1) Characteristics of the control system;
2) Limitation of energy storage systems and renewable energy resources.

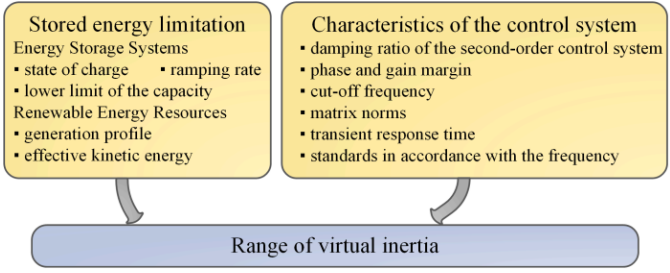

Fig. 9 Virtual inertia design framework

For each aspect, there have been some studies considering it to design the virtual inertia. However, a complete design framework hasn't been presented. Existing design methods of the virtual inertia are reviewed below. The range of suitable virtual inertia should result from the intersection, considering both of the two aspects.

*A. Characteristics of the control system*

A classical control system structure for inverters usually consists of upper controllers and inner loop controllers. The active power loop and reactive power loop are upper controllers that achieve power regulation. Inner loop controllers, including a voltage controller and a current controller, compensate for the terminal voltage reduced by the LC filter. The virtual inertia is a control parameter of upper controllers. Generally, the inertia parameter designing and the characteristics of inner loop controllers are separated [11]. Besides, reference [39] shows the coupling effect between the active power loop and reactive power loop can be neglected by tuning the parameters. Consequently, as shown in Fig. 10, the inertia parameter design can be designed independently from the control parameters of the reactive power loop and inner loop controllers, which greatly simplify the design procedure.

Many design methods for the active power loop have been proposed. Reference [11] indicates that the damping ratio of the second-order control system should be less than one so that the VSG control system can be underdamped. Besides, the phase and gain margin of the controller should be in a specific range to obtain a desirable response [45, 46]. Reference [47] proposed a design method to guarantee stability by limiting the cut-off frequency to a restricted range. Based on some properties of matrix norms, reference [13] proposed an optimal tuning of the virtual inertia parameter. Reference [48] shows large virtual inertia leads to a large $H_2$ norm, which means greater ability to reject frequency disturbances.

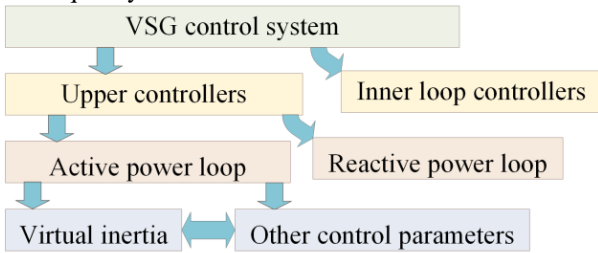

Fig. 10 Virtual inertia design considering the control system

The design of virtual inertia and damping factor cannot be separated. However, all the methods stated above can only help to determine the range of both the virtual inertia and the damping factor rather than one of them. To further determine the value of virtual inertia, some methods have been proposed. Standards in accordance with the frequency are introduced in [11] to determine the damping factor. Reference [49] establishes the optimal model of the transient response time to adjust virtual inertia and damping factor adaptively.

*B. Limitation of Energy Storage Systems and Renewable Energy Resources*

*1) Energy storage systems*

The virtual inertia doesn't provide extra energy but a greater ability to regulate the process of energy conversion of energy storage systems. The state of charge (SOC) of the energy storage systems when the VSG operates in a steady-state should be around half of the nominal capacity. And the lower and upper limits of SOC should be 20% and 90% of the maximum charge [44]. The ramping rate or response speed of different types of energy storage systems also impose limitations. The super-capacitor and flywheel can respond very fast but usually have small energy density. The hydrogen storage and electrochemical battery, contrarily, act slowly, but the energy density is higher. Hence, reference [53] proposed a hybrid control using super-capacitors to depress power fluctuations quickly and batteries for the long-term disturbances.

Reference [52] states that the stored energy and the kinetic energy of the VSG using a storage capacitor/battery should be equivalent when the system operates in a steady-state. According to the models in [25], the lower limit of the capacity is obtained for a VSG given the range of the power disturbance. Consider the VSG is subject to a disturbance $\Delta P^*$. The lower limit capacity of the energy storage system with a different damping ratio $\varsigma$ of the control system is given in Table IV.

TABLE IV
LOWER LIMIT CAPACITY OF THE ENERGY STORAGE SYSTEM WITH $\Delta P^*$

| damping ratio | lower limit capacity |
|---|---|
| $0 < \varsigma < 1$ | $\dfrac{\Delta P^*}{\omega_n \sqrt{1-\varsigma^2}} \left( \sin(2\theta_d) - e^{-\dfrac{\varsigma(\pi-\theta_d)}{\sqrt{1-\varsigma^2}}} \sin(\pi + \theta_d) \right)$ |
| $\varsigma = 1$ | $\dfrac{2\Delta P^*}{\omega_n}$ |
| $\varsigma > 1$ | $\dfrac{2\varsigma \Delta P^*}{\omega_n}$ |

Here $\varsigma = 0.5 k \sqrt{1/(\omega_{ref} H \lambda)}$, $\omega_n = \sqrt{\omega_s \lambda / H}$, $\theta_d = a\tan\left(\sqrt{1-\varsigma^2}/\varsigma\right)$.

*2) Renewable energy resources*

A PV system usually adopts a maximum power point tracking (MPPT) control to operate at the maximum power point, and hence produces an output voltage that is not constant.

Reference [36] indicates the inertia emulation through PV generation should consider the PV power generation setpoint from the MPPT controller, the rate of change of active power generation, and the energy stored in the dc-link capacitor. Reference [24] proposed an inertia design method by investigating the required power and energy of the PV system. The inertia is determined according to the maximum of [$\Delta p$, $\Delta P_f \Delta f$]. Here, $\Delta p$ is the power disturbance from PV or loads, and it lies in some prior knowledge of the PV intermittence. $\Delta f$ is the grid frequency step in a specific grid condition. The expression $\Delta P_f$ is given below, where $G_{P\omega}$ is the transfer function between active power output and angular frequency. $\Delta P_f$ denotes the maximum power amplitude when grid frequency changes 1 Hz.

$$\Delta P_f = 2\pi |G_{P\omega}|_{\max} \tag{14}$$

For VSG using kinetic energy as inertia sources, the inertia will reach its maximum value when all kinetic energy of the rotating equipment is released. In [51], a simulative wind generator set based on the VSG with a capacity of 20kVA is designed. In [54], the inertia is set according to the releasable kinetic energy of the doubly-fed induction generator (DFIG) and the minimum value of the power and torque limitations. Reference [28] uses the "effective kinetic energy" $\Delta E_k$ of the DFIG, the summary of the kinetic energy of the DFIG $\Delta E_D$ and energy fluctuation $\Delta E_P$, to determine the virtual inertia. The "effective kinetic energy" and the virtual inertia are given as below.

$$\begin{cases} H = \dfrac{\Delta E_k}{S_n} \\ \Delta E_k = \underbrace{\alpha_D (\omega_0^2 - \omega_{\min}^2)}_{\Delta E_D} + \underbrace{\int_{t_{on}}^{t_{off}} (P_m(t) - P_m(t_0))}_{\Delta E_P} \end{cases} \tag{15}$$

Here $S_n$ is the rated capacity of the DFIG, $\alpha_D$ which is a constant related to the configuration. $P_m(t)$ is the mechanical input.

Apart from the two aspects stated above, inverter capacity and PLL accuracy (for current-source VSGs) also impose limits on the selection of virtual inertia [19]. They are not described in detail here to avoid repetition, because these aspects impose limitations on all inverters regardless of the control scheme.

## VI. ADAPTIVE INERTIA CONTROL STRATEGIES

Section V discusses how to determine the range of virtual inertia for a VSG with fixed inertia. As indicated above, the range of virtual inertia is much more extensive than that of rotating inertia of a real synchronous generator. The virtual inertia control is much more flexible, and the parameter could be adjusted adaptively over time. For example, a controller adopting deviation-dependent inertia [50], an adaptive-parameters inertia control strategy [54], and a disturbance-adaptive virtual inertia control scheme [55] have been proposed for doubly fed induction generators to support the frequency control of the power system. Based on the fuzzy control theory, an adaptive virtual inertia control strategy of the wind-fire system is proposed in [56] to strengthen wind generators' ability frequency regulation. These studies are all about inertia control strategies using the kinetic energy of rotating equipment as the source of inertia emulation. Only short-term frequency regulation can be provided.

On the other hand, the virtual inertia provided by energy storage devices is continuous and flexible enough to allow different control strategies to meet the operational requirements, which shows much more potential [60]. Many researchers have reported on adaptive inertia control strategies using energy storage devices [11, 13, 17, 32, 45, 49, 52, 57], and a sample of them are discussed below.

### A. Stabilization Strategy with Fast Damping

As stated in section IV-A, the virtual inertia influences the responding speed of the VSG controller. Specifically, virtual inertia has a reverse relation to the rate of acceleration or deceleration. Reference [32] compares the stability of the VSG to a real synchronous generator by analyzing the power-angle relationship. Based on Bang-bang control, an inertia control strategy is then proposed.

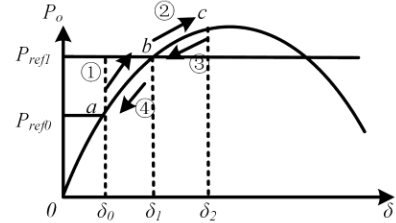

Fig. 11 Reverse relation to the rate of acceleration or deceleration

As shown in Fig.11, when the power input changes from $P_{ref0}$ to $P_{ref1}$ abruptly due to some disturbances, the operation point will move from point a to c along the power-angle curve and oscillate before finally reaching point b, the new equilibrium point. One cycle of the oscillation consists of four segments. The related condition is summarized in Table V. To obtain a damped response of the VSG, reference [32] proposes that large inertia is adopted during acceleration phases (a to b and c to b) to reduce the acceleration and small inertia during deceleration phases (b to c and b to a) to boost the deceleration.

Based on these mechanisms, reference [57] goes further by proposing frequency stabilization methods with adaptive inertia adjusted continuously. The simulation results show that the proposed strategy could enhance the damping effect of a VSG under oscillations. A mode-adaptive power-angle control method of the VSG is proposed in [27] to improve transient stability. Apart from $\Delta\omega$ and $\Delta\omega'$, the sign of the state variables $\Delta P_o$ is also detected so that the VSG could switch to more different modes.

TABLE V

FOUR SEGMENTS OF THE OSCILLATION

| Segment | $\Delta\omega$ | $\Delta\omega'$ | Mode | Alternating inertia |
|---|---|---|---|---|
| a→b | $\Delta\omega > 0$ | $\Delta\omega' > 0$ | Accelerating | Big value |
| b→c | $\Delta\omega > 0$ | $\Delta\omega' < 0$ | Decelerating | Small value |
| c→b | $\Delta\omega < 0$ | $\Delta\omega' > 0$ | Accelerating | Big value |
| b→a | $\Delta\omega < 0$ | $\Delta\omega' < 0$ | Decelerating | Small value |

### B. Dual-adaptivity Control Considering Frequency and Power Regulation

As Table III shows, the virtual inertia introduces contradictions to power and frequency regulation. However, many control strategies only consider frequency regulation ignoring power regulation. An inertia control algorism with dual-adaptivity has been proposed in [11] to address the problems, whose governing equation is given below, and the control curve is shown in Fig.12.

$$\begin{cases} H = \dfrac{H_h k_a^2 \left(\dfrac{\omega-\omega_{ref}}{\omega_{ref}}\right)^2 + H_0}{k_a^2 \left(\dfrac{\omega-\omega_{ref}}{\omega_{ref}}\right)^2 + 1} \\ \\ k_a^2 = k_g \dfrac{\left(\dfrac{\omega-\omega_{ref}}{\omega_{ref}}\right)^2}{\left(\dfrac{\omega-\omega_{ref}}{\omega_{ref}}\right)^2 + \left(\dfrac{P_o-P_{ref}}{P_{ref}}\right)^2 + 1} \end{cases} \quad (16)$$

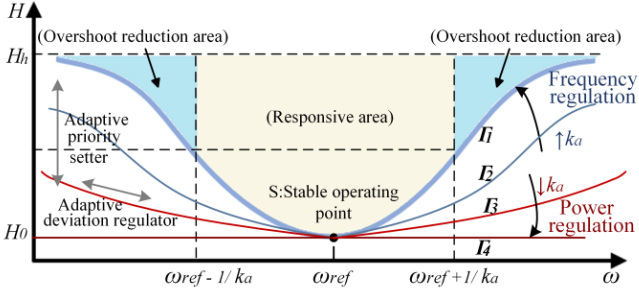

Fig. 12 Dual-adaptivity inertia control (Source: [11])

Here $H_0$ is the lower limit of virtual inertia and $H_h$ is the upper limit. $k_a$ is a sensitivity factor.

When the sensitivity factor indicates that the frequency deviation is large, the inertia is adjusted dynamically according to the frequency deviation. The systems will operate in the responsive area when the deviation is not large. Once the frequency deviation becomes too large, inertia will increase to damp oscillations. This is the first adaptivity.

When the sensitivity factor indicates that the power deviation is large, the inertia decreases. With small inertia, the overshoot is slight, and the resettling time is short, according to Table III. This is the second adaptivity. Therefore, the proposed strategy could achieve fast response without compromising small overshoot, and it also achieves frequency regulation without compromising power regulation.

### C. Optimal Control of Transient Response

The studies discussed above offer solutions caused by fixed-inertia control, but they separate the design of virtual inertia and other control parameters. Reference [49] proposes a VSG control strategy where the virtual inertia is adjusted along with the damping factor by establishing the optimal model to find the minimum transient response time $T$ as below.

$$\begin{cases} \min_{\omega} T \\ s.t. \ \Delta\delta = \int_0^T (\omega(t)-\omega_0)dt \ , \omega(t)|_{t=0} = \omega(t)|_{t=T} = \omega_0, \\ |\omega(t)-\omega_0| \leq 2\pi\Delta f_{max} \ , \omega(t)' \leq 2\pi\kappa \end{cases} \quad (17)$$

where $\Delta\delta$ is the angle deviation under disturbances. $\Delta f_{max}$ and $\kappa$ are the threshold of frequency deviation and change rate of angular frequency.

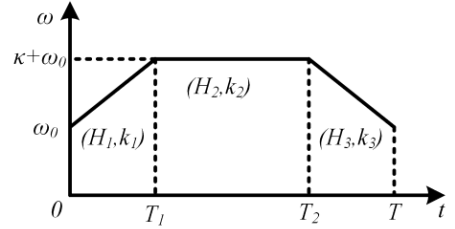

Fig. 13 The response curve of angular frequency

As shown in Fig.13, the response curve of angular frequency is divided into three processes to obtain the shortest response time. According to the state equation of each process, the virtual inertia and damping factor in each process are as below. A zero damping factor is adopted in process 1 and 3 to speed up the transient response. The proposed strategy alleviates the fluctuation of frequency. However, the method relies on a detector to determine in which process the VSG controller is.

$$\begin{cases} H_1 = \dfrac{P_m-P_e}{2\pi\kappa\omega}, \ H_2 = H_0 \dfrac{P_m-P_e}{2\pi k \omega}, \ H_3 = -\dfrac{P_m-P_e}{2\pi\kappa\omega}, \\ k_1 = 0, \ k_2 = \dfrac{P_m-P_e}{2\pi\Delta f_{max}(\omega_0+\Delta f_{max})}, \ k_3 = 0 \end{cases} \quad (18)$$

## VII. RECOMMENDATIONS FOR FURTHER RESEARCH

1. For most existing adaptive control strategies of a VSG, only one parameter (virtual inertia or damping factor) is designed as adaptive, but the other parameter is fixed. The study discussed in section VI-C only focuses on the responding time but ignored other dynamic characteristics. Therefore, a control strategy where several control parameters (including virtual inertia) are adjustable dynamically could be proposed to improve a VSG's performance further. Considering that it is challenging to design several parameters at the same time just by control theory or storage limitation, optimization theory might be a useful tool. By introducing a proper optimization model, different aspects of dynamic characteristics (overshot and responding speed, etc.), stability margin, storage limitation, and so on could be considered to achieve a multi-objective optimal design of virtual inertia control.

2. Nowadays, distributed generators like photovoltaic and wind generators usually appears in the form of hybrid systems, where an energy storage system is equipped. The problem of virtual inertia configuration for these hybrid systems arises. For example, how to assess the energy source when both kinetic energy and storage energy exist in one system.

3. Multiple VSGs may adopt a different value of inertia, or there might be some cooperative inertia control strategies. Therefore, how to design inertia for a system with multiple VSGs remains to be studied.

VIII. CONCLUSION

Virtual inertia provides an inertia property that distinguishes the VSG from other control schemes. In this paper, the research framework of virtual inertia control is established by providing a comprehensive review. First, according to the definition, virtual inertia is an integration constant of the VSG controller. Then, the sources of inertia emulation and their features are discussed according to the type of energy. The influencing mechanisms of the virtual inertia are discussed based on different aspects of the VSG's performance, including dynamic characteristics, stability margin, and transient energy. The design method of the virtual inertia is presented considering both the characteristics of the control system and the limitation of energy storage systems and renewable energy resources. Several adaptive inertia control strategies are reviewed to show how to control inertia to meet different operation requirements. Finally, some suggestions for further researches are presented to help the understanding and application of virtual inertia control.

MEIYI LI (S'17-M'20) received the B.Sc. and M.Sc. degree in electrical engineering from Shanghai Jiao Tong University (SJTU), Shanghai, China, in 2017 and 2020, respectively. She is now pursuing a Ph. D. degree in Department of Electrical and Computer Engineering, Carnegie Mellon University. Her research interests include stability and control of distributed generation.

WENTAO HUANG(S'15–M'16-SM'20) was born in Anhui, China, in 1988. He received the Ph.D. degree in electrical engineering from Shanghai Jiao Tong University, Shanghai, China, in 2015. He is currently an Associate Professor with the Department of Power Electrical Engineering. His research interests include protection and control of active distribution systems, microgrids, smart grid, and renewable energy.

NENGLING TAI(M'07) received the B.Sc., M.Sc., and Ph.D. degrees in electrical engineering from the Huazhong University of Science and Technology (HUST), Wuhan, China, in 1994,1997, and 2000, respectively. He is currently a Professor with the Department of Power Electrical Engineering, Shanghai Jiao Tong University, Shanghai, China. His research interests include the protection and control of active distribution systems, microgrids, smart grid, and renewable energy.